\documentstyle[psfig,referee]{laa}
\thesaurus{12.03.4; 12.12.1; 11.06.1}
\title{The effect of non-radial motions on the X-ray temperature distribution
function and the two-point correlation function of clusters.}
\author{A. Del Popolo\inst{1,2}, and ~ M. Gambera\inst{1,3}}
\offprints{M. Gambera}
\institute{$^1$ Istituto di Astronomia dell'Universit\`a di Catania, 
Viale A.Doria, 6 - I 95125 Catania, ITALY \\
$^2$ Facolt\`{a} di Ingegeneria, Universit\`{a} Statale di Bergamo, Piazza Rosate, 2 - I 24129 Bergamo, ITALY \\
$^3$ Osservatorio Astrofisico di Catania and CNR-GNA, 
Viale A.Doria, 6 - I 95125 Catania, ITALY 
}
\date{}
\begin{document}
\maketitle

\begin{abstract}
We show how non-radial motions, originating
in the outskirts of clusters of galaxies, may
reduce the discrepancy between the Cold Dark Matter (CDM) predicted
X-ray temperature distribution function
of clusters of galaxies and the observed one and
also the discrepancy between the CDM
predicted two-point correlation function of clusters of galaxies and that
observed. 
We compare
Edge et al. (1990) and Henry \& Arnaud (1991) data with 
the distribution function of X-ray temperature, 
calculated using
Press- Schechter's (1974 - hereafter PS) theory and 
Evrard's (1990) prescriptions for the mass-temperature 
relation and taking account of the non-radial motions
originating from the gravitational interaction of the quadrupole moment
of the protocluster with the tidal field of the matter of the
neighboring protostructures. 
We find that the model produces a reasonable 
clusters temperature distribution.
We compare the two-point cluster correlation function which
takes account of the non-radial motions  both with that obtained
by Sutherland \& Efstathiou (1991), from the analysis
of Huchra's et al. (1990) deep redshift survey, and with the
data points for the Automatic Plate Measuring (APM) clusters, computed 
by Efstathiou et al. (1992a),
showing that non-radial motions reduce the discrepancy between
the theoretical and the obsservational correlation function.

\end{abstract}

\section{Introduction}

Although at the begining the standard form
of CDM was very successful in describing the structures observed in the
Universe (galaxy clustering statistics, structure formation epochs, peculiar
velocity flows) (Peebles, 1982; Blumenthal et al. 1984; Bardeen et al. 1986 -
hereafter BBKS; 
White et al. 1987; Frenk et al. 1988; Efstathiou 1990) recent measurements
have shown several deficiencies in the model, at least, when any bias of the
distribution of galaxies relative to the mass is constant with scale
(see Babul \& White 1991; Bower 1993; Del Popolo \& Gambera 1998a, 1998b). Some
of the most difficult problems that must be reconciled  with the theory are:
\begin{description}
\item the magnitude of the
dipole of the angular distribution of optically selected galaxies (Lahav et
al. 1988; Kaiser \& Lahav 1989);
\item the possible observations of clusters of
galaxies with high velocity dispersion at $z\geq 0.5$ (Evrard 1989);
\item the
strong
clustering of rich clusters of galaxies, 
$ \xi_{cc}(r) \simeq (r / 25h^{-1}Mpc)^{-2}$, 
far in excess of CDM predictions (Bahcall \&
Soneira 1983);
\item the X-ray temperature distribution function of clusters,
over-producing the observed clusters abundances (Bartlett \& Silk 1993); 
\item the
conflict between the normalization of the spectrum of the perturbation which
is required by different types of observations;
\item 
the incorrect scale dependence of the galaxy correlation
function, $\xi (r)$, on scales $10$ to $100$ $h^{-1} Mpc$, having $\xi (r)$ too
little power on the large scales compared to the power on smaller scales
(Maddox et al. 1990a; Saunders et al. 1991; Lahav et al. 1989;
Peacock 1991; Peacock \& Nicholson 1991).
\end{description}
These discrepancies between the theoretical predictions of the CDM model
and  the observations led many authors to conclude that the shape of the
CDM spectrum is incorrect and to search alternative models
(Peebles 1984; Efstathiou et al. 1990a; Turner 1991; White et al. 1993a, 1993b;
Shafi \& Stecker 1984; Valdarnini \& Bonometto 1985; 
Holtzman 1989;
Schaefer 1991; Shaefer \& Shafi 1993; Holtzman \& Primack 1993;
Cen et al. 1992; Bower et al. 1993).\\
In this paper we address two of the quoted and most serious
problems of the CDM model,
namely that of the 
discrepancy between the predicted and observed X-ray
temperature distribution function of clusters and that of 
the discrepancy between 
the CDM predicted two-point correlation function of clusters of galaxies and 
that observed, 
showing how them
may be reduced when
non-radial motions, that develop during the collapse process,
are taken into account.\\
X-ray studies of clusters of galaxies have provided a great number
of quantitative data for the study of cosmology. The mass of a rich
cluster is approximately $10^{15} h^{-1} M_{\odot}$, where
$h= H_0/(100 km s^{-1} Mpc^{-1})$, being
$H_0$ the Hubble constant at current epoch (in the paper we adopt $h=1/2$).
This mass comes from a region of diameter
$\simeq 20 h^{-1} Mpc$ and consequently the observations of clusters can
provide information on the mass distribution of the Universe on these
scales. Furthermore, since rich clusters  are rare objects, their properties
are expected to be sensitive to the underlying mass density
field from which they arose. On scales of $\simeq 20 h^{-1} Mpc$
the density field
is still described by the linear perturbation theory so the  measurement at
the present epoch can be related to  the initial conditions.
While the gravitational mass in bound objects is easily computed using
analytic approaches such as the PS formalism 
or N-body simulations, it is not so easily measured
The most robust measurements of the clusters
abundance currently rely on the clusters temperature function, that is
the
number density of clusters above  a certain temperature $N(>kT)$ expressed
in units of $h^3$ Mpc$^{-3}$.
The reasons of this choice are several. The integrated temperature does not
suffer from projection effects, and complete samples of clusters may be
obtained from all sky X-ray surveys. 
Clusters are close to
isothermal, both observationally and in simulations, which makes their
temperature determination robust and insensitive to numerical
resolution or  telescope angular resolution. The temperature of a
cluster depends primarily on the depth of the potential well of the
dark matter, and the state of equilibrium of the gas. This is in
contrast to the clusters luminosity, which depends strongly on small
scale parameters like clumping and core radius.\\
The clusters abundance is one of the best observables available for
determining the density field. 
In fact, according to the gravitational instability scenario, galaxies and clusters
form where the density contrast, $\delta$, is large enough so that the surrounding
matter can separate from the general expansion and collapse. Consequently
the abundance of collapsed objects depends on the amplitude of the
density perturbations. In the CDM model these latter follow a gaussian
probability distribution and their amplitude on a scale $R$
is defined by $\sigma(R)$, the r.m.s. value of $\delta$,
which is related to the power
spectrum, $P(k)$. In hierarchical models of structure formation, like
CDM, $\sigma(R)$ decreases with increasing scale, $R$, and consequently
the density contrast required to form large objects, like clusters of galaxies,
 rarely occurs. The present abundance of clusters is then extremely
sensitive to small change in the spectrum, $P(k)$. Moreover the rate
of clusters evolution is strictly connected to the density parameter, $\Omega_0$.
Then clusters abundance and its evolution are a probe of $\Omega_0$ and $P(k)$
and can be used to put some constraints on them.   \\
Henry and Arnaud (1991)
and Edge et al. (1990) have analysed the HEAO-1 A2
all sky X-ray survey to obtain temperatures for clusters at a flux
limit of $F_{2 \div 10\rm keV} > 3\times 10^{-11}$ erg/(cm$^2$ sec).  The
sample contains the 25 brightest X-ray  clusters and is  complete for more than   90 \, \% at galactic latitude $|b|>20^o$.  By virtue of their
brightness the temperatures are easily and well determined. They also
constructed temperature distributions that has been compared with
the CDM predicted X-ray temperature distribution function
(Bartlett \& Silk 1993). Using this data it was shown that galaxy
clusters are too numerous in the CDM scenario (Bartlett \& Silk 1993).\\
The other problem of the CDM model that we are addressing  is that of the discrepancy between 
the two-point correlation function of clusters of galaxies and the observed
one. \\
Measurements of galaxy clustering on large scales (Maddox et al 1990a;
Efstathiou et al. 1990b; Saunders et al. 1991) revealed rms fluctuations of
the order of 50\% within spheres of radius 20$h^{-1}Mpc$. These amplitudes,
relative to non-linear clustering on scales 5$h^{-1}Mpc$, are 2-3 times
larger than that predicted by the CDM model. At first the suggestion of 
excess power arose from the estimates of the autocorrelation
of Abell clusters (Hauser \& Peebles 1973; Bahcall \& Soneira 1983;
Klypin \& Kopylov 1983). As shown by White et al. (1987), the cluster
correlations in {\it Standard} Cold Dark Matter (SCDM) should 
have an amplitude 2-3 times smaller than these estimates.
Although the integrity of Abell cluster has been repeatedly called into
question (Sutherland 1988; Dekel et al. 1989; Olivier et al. 1990; Sutherland
\& Efstathiou 1991) and a new sample of rich clusters identified from the
APM survey exhibits weaker correlations, marginally consistent with the CDM
predictions (Dalton et al. 1992),
X-ray clusters samples (Lahav et al. 1989) have a large correlation length
and observations of bright radio galaxies (Peacock 1991; Peacock \&
Nicholson 1991) are also strongly clustered on large scales.\\
Finally, as shown in some studies of galaxy clustering on large              
scales (Maddox et al. 1990a; Efstathiou et al. 1990b; Saunders et al. 1991),
the measured rms fluctuations within spheres of radius $20h^{-1}Mpc$ have
values 2-3 times larger than that predicted by the CDM model.
In order to solve the quoted discrepancies between CDM model previsions
and observations, alternative models have been introduced.
Several authors (Peebles 1984;
Efstathiou et al. 1990a; Turner 1991) have lowered 
the matter density under the critical value ($%
\Omega _m<1$) and have added a cosmological constant in order to retain
a flat Universe ($\Omega _m+\Omega _\Lambda =1$). The spectrum of the
matter density is specified by the transfer function, but its shape is
affected because the epoch of matter-radiation equality
(characterized by the redshift $z_{eq}$) is
earlier, $1+z_{eq}$ being increased by a factor $1/\Omega _m$. 
Around the epoch of redshift $z_\Lambda $,
where $z_{\Lambda}=(\Omega_{\Lambda}/\Omega_m)^{1/3}-1$,
the effect of the cosmological constant becomes
important and the growth of the density contrast slows down
and ceases after $z_\Lambda $. As a consequence, the normalisation of the
transfer function begins to fall, even if its shape is retained. 
In particular Bartlett \& Silk (1993)
showed that a model with $\Omega_0=0.2$ and $\Omega_\Lambda=0.8$ produces
a reasonable temperature distribution of clusters, but a higher normalization with
$\sigma_8 \simeq 2$, 
(where
$\sigma_8$ is the
rms value of $\frac{\delta M}{M}$ in a sphere of $8 h^{-1}$Mpc),
is needed to explain the peculiar velocity field
(Efstathiou et al. 1992b). \\
Mixed dark matter models (MDM) (Shafi \& Stecker 1984;
Valdarnini \& Bonometto 1985; Schaefer et al. 1989; Holtzman 1989; Schaefer
1991; Schaefer \& Shafi 1993; Holtzman \& Primack 1993) increase the
large-scale power because neutrinos free-streaming damps the power on small
scales. In particular the model with $\Omega_{HDM}=0.3$ simulated by
Davis et al (1992), gives a good reproduction
of the clusters abundance with a normalization within the 1 $\sigma$
errors of the COBE. These last models have some difficulty in reproducing
peculiar velocities (Efstathiou et al. 1992b).\\
Alternatively, changing the primeval spectrum, several problems of CDM
are solved (Cen et al. 1992). 
Finally 
it is possible
to assume that the
threshold for galaxy formation
is not spatially invariant but weakly modulated ($2\%-3\%$ on scales $
r>10h^{-1}Mpc$) by large scale density fluctuations, with the result that
the clustering on large-scale is significantly increased (Bower et al.
1993). This model follows the spirit of the well known
high-peak model but differs from it because nonlocal
physical processes produce different shapes of the mass and galaxy
correlation function.\\  
In any case the solution to the quoted problems 
till now proposed is related to alternative models with
more large-scale power than CDM.\\
Here, we propose a solution  to the problem using the CDM model and
taking account of the non-radial motions
originating from the gravitational interaction of the quadrupole moment
of the protocluster with the tidal field of the matter of the
neighboring protostructures.\\
The plan of this work is the following: in Sect. ~2 we introduce 
the model used to find the effects 
of non-radial motions 
on the X-ray temperature distribution function and the two-point 
correlation function. In Sect. ~3 we compare
the results of this model with the X-ray
temperature distributions given by Henry \& Arnaud (1991) and 
Edge et al. (1990) while in Sect. ~ 4 we compare
our model with the
two-point correlation function obtained by  
Sutherland \& Efstathiou (1991), from the analysis
of Huchra's et al. (1990) deep redshift survey, and with the
data points for the Automatic Plate Measuring (APM) clusters, computed 
by Efstathiou et al. (1992a).
Sect. ~5 is devoted
to the conlusions.\\

\section{The X-ray temperature function}

The PS theory provides an analytical description of the
evolution of structures in a hierarchical Universe. In this model the linear
density field, $\rho ({\bf x},t)$, is an isotropic random Gaussian field, the
non-linear clumps are identified as over-densities (having a density contrast 
$\delta _c\sim 1.68$ - Gunn \& Gott 1972) in the linear density field, while
a mass element is incorporated into a non-linear object of mass $M$ when the
density field smoothed with a top-hat filter of radius $R_f$, exceeds a
threshold $\delta _c$ ($ M\propto R_f^3$).
The comoving number density of non-linear objects of mass $M$ to $M+dM 
$ is given simply by:
\begin{equation}
N(M,t)dM=-\rho_b \sqrt{\frac 2\pi }\nu \exp \left( -\nu ^2/2\right) \frac
1\sigma \left( \frac{d\sigma }{dM}\right) \frac{dM}M  
\label{eq:press}
\end{equation}
where $\rho_b$ is the mean mass density, $\sigma (M)$ is the rms linear mass
overdensity
evaluated at the epoch when the mass function is desidered and $\nu =%
\frac{\delta _c}{\sigma (M)}$. The redshift dependence of
Eq. ~(\ref{eq:press}) can be
obtained remembering that
\begin{equation}
\nu =\frac{\delta _c(z)D(0)}{\sigma _o(M)D(z)}
\end{equation}
being $ D(z)$ the growth factor of the density perturbation and
$\sigma_o(M)$ the current value of $\sigma (M)$.
In Eq. ~(\ref{eq:press})
PS introduced arbitrarily a factor of two because 
$ \int_0^\infty dF(M)=1/2$, so that only half of the mass in the Universe
is accounted for. Bond et al. (1991) showed that the "fudge factor" 2 is
naturally obtained using the excursion set formalism in the sharp
$k$-space while for general filters (e.g., Gaussian or "top hat")
it is not possible to obtain an analogous analytical result. As stressed by
Yano et al. 1996, the factor of 2 obtained in the sharp $k$-space
is correct only if the spatial correlation of the density fluctuations
is neglected.
In spite of the quoted problem, several authors (Efstathiou et al.
1988; Brainerd \& Villumsen 1992; Lacey \& Cole 1994) showed that PS analytic
theory correctly agrees with N-body simulations.
In particular Efstathiou et al. (1988),
showed
that PS theory correctly agrees with                 
the evolution of the distribution of mass amongst groups and clusters of
galaxies (multiplicity function). Brainerd \& Villumsen (1992) studied the
CDM halo mass function using a hierarchical particle mesh code. From this
last work  it results that PS formula fits the results of the simulation up
to a mass of ~10 times the characteristic 1$\sigma$ fluctuation mass,
$M_{\ast}$, being $M_{\ast} \simeq 10^{15} b^{-6/(n_{l}+3)} M_{\odot}$,
where $b$ is the
bias parameter 
and $n_{l}$ is the local slope of the power spectrum.
PS theory has proven particularly useful in
analyzing the number counts and redshift distributions for QSOs (Efstathiou
\& Rees 1988), Lyman $\alpha $ clouds (Bond et al. 1988) and
X-ray clusters (Cavaliere \& Colafrancesco 1988). \\
Some difficulties arise when PS theory is compared with observed
distributions. To estimate the multiplicity function of real systems  one needs to know the temperature-mass (T-M) relation in
order to trasform the mass distribution into the temperature distribution.
Theoretical uncertainty arises in this transformation because the exact
relation between the mass
appearing in the PS expression and the temperature of the intracluster
gas is unknown.
Under the standard assumption of the 
Intra-Cluster (IC) gas in hydrostatic equilibrium with the potential well 
of a spherically simmetric, virialized cluster, the IC gas temperature-mass
relation is easily obtained by applying the virial
theorem
and for a flat matter-dominated Universe
we have that (Evrard 1990, Evrard et al. 1996, Evrard 1997, Bartlett 1997):
\begin{equation}
T=(6.4h^{2/3}keV)\left( \frac M{10^{15}M_{\odot }}\right) ^{2/3}(1+z) 
\label{eq:ma2}
\end{equation}
The assumptions of perfect hydrostatic equilibrium and virialization
are in reality not completely satisfied in the case of clusters. Clusters profile may
depart from isothermality, with sligth temperature gradients
throughout the cluster. The X-ray weighted temperature can be slightly
different from the mean mass weighted virial temperature. In any
case
the scatter in the T-M relation given by Eq. (\ref{eq:ma2})
is of the order of 
$\simeq 10 \%$ (Evrard 1991).
The mass variance present in Eq.~(\ref{eq:press})
can be obtained once a spectrum, $P(k)$, is fixed:
\begin{equation}
\sigma ^2(M)=\frac 1{2\pi ^2}\int_0^\infty dkk^2P(k)W^2(kR) 
\label{eq:ma3}
\end{equation}
where $W(kR)$ is a top-hat smoothing function:
\begin{equation}
W(kR)=\frac 3{\left( kR\right) ^3}\left( \sin kR-kR\cos kR\right) 
\label{eq:ma4}
\end{equation}
and the power spectrum $P(k)=Ak^nT^2(k)$ is fixed giving the transfer
function $T(k):$
\begin{eqnarray}
T(k) &=& [\ln \left( 1+4.164k\right)]^2 \cdot (192.9+1340k+ \nonumber \\
& + &  1.599\cdot 10^5k^2+1.78\cdot 10^5k^3+3.995\cdot
10^6k^4)^{-1/2}
\label{eq:ma5}
\end{eqnarray}
(Ryden \& Gunn 1987; BBKS)
and $ A$ is the normalizing constant.
The normalization of the
spectrum may be obtained in several ways. One possibility is to normalize
it  to
COBE scales using the cosmic microwave anisotropy quadrupole
$Q_{rms-PS}=17 \mu K $. This corresponds to $\sigma_8=0.95 \pm 0.2$ 
(Smoot et al. 1992; Liddle \& Lyth 1993). More recent determinations give 
$\sigma_8=1.33$ if we use
the BBKS spectrum, while $\sigma_8=1.22$ if we use the
spectrum by Bond \& Efstathiou (1984) (Klypin et al. 1997).
Another way of fixing the normalization is via the abundance of clusters
giving $\sigma_8= 0.5 \div 0.6$ (Pen, 1997; Oukbir 1996;
Bartlett 1997). Normalisation on scales from $10$ to $50Mpc$ obtained from QDOT
(Kaiser et al. 1991) and POTENT (Dekel et al. 1992) 
requires that $\sigma _8$
is in the range $0.7 \div 1.1$, which is compatible with COBE normalisation
while the observations of the pairwise velocity dispersion of galaxies on
scales $r\leq 3Mpc$ seem to require $\sigma _8<0.5$. Our normalization,
$\sigma_8=1$,
is intermediate between that suggested by clusters abundance and that of
COBE.\\
As shown by Bartlett \& Silk (1993) the X-ray distribution
function obtained using a standard CDM spectrum over-produces the clusters
abundances data obtained  by Henry \& Arnaud (1991) and Edge et al.
(1990). The
discrepancy can be reduced taking into account the non-radial motions that
originate when a cluster reaches the non-linear regime.
In fact, the PS temperature distribution requires the specification
of $\delta _c$ and the temperature-mass relation T-M. The presence of non-radial
motions changes both $\delta _c$ and the T-M relation. Barrow \& Silk
(1981), Szalay \& Silk 1983 and Peebles 1990 assumed that non-radial motions
would be expected (within a developing protocluster) due to the tidal
interaction of the irregular mass distribution around the protocluster
with the
neighbouring protoclusters. The kinetic energy of these non-radial motions
inhibits the collapse of the protocluster enabling  it to reach
statistical equilibrium before the final collapse (Davis \& Peebles 1977;
Peebles 1990). The role of non-radial motions has been also pointed out by
Antonuccio \& Colafrancesco (1995). After deriving the conditional
probability distribution $f_{pk}(\mathbf{v}|\nu )$ of the peculiar velocity
around a peak of a Gaussian density fiely non-radial motions.
In these regions the fate of the infalling material could be influenced by
the amount of tangential velocity relative to the radial one. \\ 
This can be shown
writing the equation of motion of a spherically symmetric mass distribution
with density $n(r)$:
\begin{equation}
\frac \partial {\partial t}n \langle v_r \rangle +\frac \partial {\partial r}n 
\langle v_r^2 \rangle + \left(2 \langle v_r^2 \rangle - 
\langle v_\vartheta ^2 \rangle \right) \frac nr+n(r)\frac \partial {\partial
t} \langle v_r \rangle = 0  
\label{eq:peeb}
\end{equation}
where $ \langle v_r \rangle$ and $ \langle v_\vartheta \rangle $ 
are, respectively, the mean radial and
tangential streaming velocity. Eq. (\ref{eq:peeb}) shows that high
tangential velocity dispersion 
$(\langle v_\vartheta ^2 \rangle \geq 2 \langle v_r^2 \rangle)$ 
may alter the infall pattern. The expected delay in the collapse of a 
perturbation may be calculated using a model due to Peebles (Peebles 1993).\\
 Let's consider an ensemble of gravitationally growing mass concentrations 
 and suppose that the material in each system collects within the
same potential well
with inward pointing acceleration given by $g(r)$. We
indicate with $dP=f(L,r v_r,t)dL dv_r dr$ the probability that a particle
can be found  in the proper radius range $r$, $r+dr$, in the radial
velocity range $v_r={\dot r}$, $v_r+d v_r$ and with angular momentum
$L=r v_\theta$ in the range $dL$.
The radial
acceleration of the particle is: 
\begin{equation}
\frac{dv_r}{dt}=\frac{L^2(r,\nu )}{M^2r^3}-g(r)  
\label{eq:coll}
\end{equation}
Eq. (\ref{eq:coll}) can be derived from a potential
and then from Liouville's theorem it follows that
the distribution function, $f$,
satisfies the collisionless Boltzmann equation:
\begin{equation}
\frac{\partial f}{\partial t} + v_{r}
\frac{\partial f}{\partial r} + \frac{\partial f}{\partial v_{r}}
\cdot \left[ \frac{L_{2}}{r^{3}} - g(r) \right] = 0
\end{equation}
Using Gunn \& Gott's (1972) notation we write the proper radius of a shell in terms of the expansion parameter, $%
a(r_i,t)$, where $r_i$ is the initial radius: 
\begin{equation}
r(r_i,t)=r_{i}a(r_i,t)
\label{eq:ma6}
\end{equation}
Remembering that $M=\frac{4\pi }3\rho (r_i,t)a^3(r_i,t)r_i^3$, 
that $\frac{%
3H_i^2}{8\pi G}=\rho_{ci}$, with $\rho_{ci}$ and $H_i$   respectively 
the critical mass density and the Hubble constant at the time $t_i$, and assuming that
no shell crossing occurs so that the total mass inside each shell remains
constant, ($\rho (r_i,t)=\frac{\rho _i(r_i,t_i)}{a^3(r_i,t)}$) Eq. (\ref
{eq:coll}) may be written as: 
\begin{equation}
\frac{d^2a}{dt^2}=-\frac{H_i^2(1+\overline{\delta })}{2a^2}+\frac{4G^2L^2}{%
H_i^4(1+\overline{\delta })^2r_i^{10}a^3}  
\label{eq:sec}
\end{equation}
where $\overline{\delta }=\frac{\rho _i-\rho_{ci}}{\rho _{ci}}$.\\
Integrating Eq. (\ref{eq:sec}) we have: 
\begin{equation}
(\frac{da}{dt})^2=H_i^2\left[ \frac{1+\overline{\delta }}a\right] +\int 
\frac{8G^2L^2}{H_i^4 r_i^{10}\left( 1+\overline{\delta }\right) ^2}\frac
1{a^3}da-2C  
\label{eq:ses}
\end{equation}
where $C$ is the binding energy of the shell. 
Integrating once more we have:
\begin{equation}
t_{ta}=\int_0^{a_{\max }}\frac{da}{\sqrt{H_i^2\left[ \frac{1+\overline{\delta} }a-
\frac{%
1+\overline{\delta }}{a_{\max }}\right] +\int_{a_{\max }}^a\frac{8G^2L^2}{%
H_i^4r_i^{10}(1+\overline{\delta} )^2}}a^3}
\label{eq:sell}
\end{equation}
Using Eqs (\ref{eq:ses}) and (\ref{eq:sell}) it is possible
to find the linear over-density at the turn-around epoch, $t_{ta}$.
In fact solving Eq. (\ref{eq:sell}),
for some epoch of interest, we may obtain the expansion parameter
of the turn-around epoch. This is related to the binding energy of the shell
containing mass $M$ by Eq. (\ref{eq:ses}) with $\frac{da}{dt}=0$.
In turn the binding energy of a growing mode solution is uniquely given
by the linear overdensity, $\delta_{i}$, at time $t_{i}$.
From this overdensity, using the linear theory, we may obtain that of
the turn-around epoch and then that of the collapse which is given by:
\begin{equation}
\delta _c(\nu )=\delta _{co}\left[ 1+\frac{8G^2}{\Omega
_o^3H_0^6r_i^{10}\overline{\delta} (1+\overline{\delta} )^2}\int_0^{a_{\max }}\frac{L^2 \cdot da}{a^3}%
\right]
\label{eq:ma7} 
\end{equation}
where $\delta _{co}=1.68$ is the critical threshold for a spherical model,
while $H_0$ and $\Omega_0$ are respectively
the Hubble constant and the density parameter at the current epoch $t_0$.
Filtering the spectrum on clusters scales, $R_f=3h^{-1}Mpc$, we obtained the
total specific angular momentum, $h(r,\nu )=L(r,\nu)/M_{sh}$, 
acquired during expansion,
integrating the torque over time (Ryden 1988 - Eq. 36): 
\begin{equation}
h(r,\nu ) = \frac{\tau_o t_0 \overline{\delta}_o^{-5/2}}{\sqrt[3]{48}M_{sh}}%
\int_0^\pi \frac{\left( 1-\cos \theta \right) ^3}{%
\left( \vartheta -\sin \vartheta \right) ^{4/3}}\frac{f_2(\vartheta) \cdot%
d\vartheta}{f_1(\vartheta )-f_2(\vartheta )\frac{\delta _o}%
{\overline{\delta _o}}}
\label{eq:ang}
\end{equation}
where $\tau_o$ $\delta_o$ and $\overline{\delta}_o$
are respectively the torque, the mean overdensity and the mean
overdensity within a sphere of radius $r$
at the current epoch $t_0$.
The functions $f_1(\vartheta )$, $f_2(\vartheta )$ are given by Ryden 
(1988 - Eq. 31):
\begin{equation}
f_1(\theta)=16-16 \cos \theta+\sin^2 \theta-9 \theta \sin \theta
\end{equation}
\begin{equation}
f_2(\theta)=12-12 \cos \theta+3 \sin^2 \theta-9 \theta \sin \theta
\end{equation}
where $\theta$ is a a parameter connected to the time, $t$,
through the following
equation:
\begin{equation}
t=\frac{3}{4} t_0 \overline{\delta}_{o}^{-3/2}(\theta -\sin \theta)
\end{equation}
The mean overdensity within a sphere of radius $r$ , $\overline{\delta }%
(r)$, is given by:
\begin{equation}
\overline{\delta} (r,\nu )=\frac 3{r^3}\int_0^r d x x^2 \delta(x)
\label{eq:ma8}
\end{equation}
The mass dependence of the threshold parameter, $ \delta_{c}(\nu)$, can be
found as follows: we calculate the binding radius, $r_{b}$, of the shell 
using Hoffmann \& Shaham's criterion (1985): 
\begin{figure}[ht]
\psfig{file=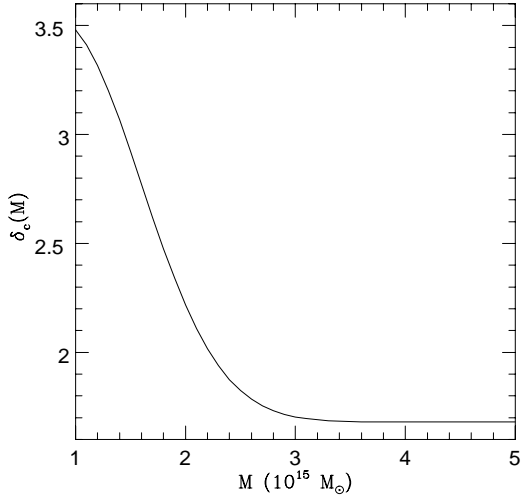,width=12cm}
\caption[]{The threshold $\delta_{c}$  as a function of the mass M, for a CDM
spectrum ($\Omega_0=1$, $h=1/2$)
with $R_{f}=3 h^{-1}Mpc$, taking account of non-radial motions.}
\end{figure}
\begin{equation}
T_{c}(r, \nu) \leq t_{0}
\label{eq:ma9}
\end{equation}
where $T_{c}(r,\nu)$ is the calculated time of collapse of a shell and $
t_{o}$ is the Hubble time. We find a relation between $ \nu$ and $M$ through
the equation $ M=4 \pi\rho_b r^{3}_{b}/3$. We so obtain $ \delta_{c}(\nu(M))$. 
In Fig. 1 we show the variation of the threshold parameter, $\delta _c(M)$,
with the mass $M$. Non-radial motions influence the value of $\delta _c$
which increases for peaks of low mass  and remaines unchanged for
high mass peaks. As a consequence, the
structure formation by low mass peaks is
inhibited. 
In other words, in agreement with the cooperative
galaxy formation theory (Bower et al. 1993), structures form more
easily in over-populated regions. As we previously told, the cooperative galaxy
formation is able to reconcile the CDM model with the APM correlations
by assuming the threshold for galaxy formation to be modulated by
large-scale density fluctuations rather than to be spatially invariant. But
there
exists some difficulty in finding a physical mechanism able to produce
the modulation. In our model this mechanism is linked to non-radial
motions.\\
To get the temperature distribution it is necessary to know 
the temperature-mass
relation. This can be obtained using the virial theorem, energy
conservation and using Eq. (\ref{eq:ses}).
From the virial theorem we may write:
\begin{equation}
\langle K \rangle = \frac{GM}{2 r_{eff}}+\int_{0}^{r_{eff}} \frac{L^2}{2 M^2 r^3} d r
\label{eq:vir}
\end{equation}
while from the energy conservation:
\begin{equation}
- \langle K \rangle +\frac{GM}{r_{eff}}+\int_{0}^{r_{eff}} \frac{L^2}{M^2 r^3} d r=
\frac{GM}{r_{ta}}+\int_{0}^{r_{ta}} \frac{L^2}{M^2 r^3} d r
\label{eq:ener}
\end{equation}
Eq. (\ref{eq:vir}) and Eq.(\ref{eq:ener}) can be solved for
$r_{eff}$ and $<K>$.
We finally have that:
\begin{equation}
T=(6.4 keV)\left( \frac{M \cdot h}{10^{15} M_{\odot }}\right)^{2/3}
\left[ 1+
\frac{\eta \psi \int_0^r \frac{L^2 dr}{M^2 r^3}}{(G^{2}
\frac{H_0^2\Omega_0}{2} M^{2})^{1/3}}\right] 
\label {eq:temp} 
\end{equation}
where
$\eta$ is a parameter given by $ \eta=r_{ta}/x_{1}$, being $r_{ta}$
the radius of the turn-around epoch, while $x_{1}$ is defined
by the relation
$M=4 \pi \rho_b x^{3}_{1}/3$ and $ \psi=r_{eff}/r_{ta}$ where $r_{eff}$
is the time-averaged radius of a mass shell.
Eq. (\ref{eq:temp}) was normalised to agree with Evrard's (1990) simulations
for $L=0$. \\

\section{Non radial motions and the X-ray temperature function.}

The new T-M relation is Eq. (\ref{eq:temp}) which differs from 
Eq. (\ref{eq:ma2}) for the presence of the term:
\begin{equation} 
\frac{\eta \psi \int_0^r \frac{L^2dr}{M^2r^3}}{(G^{2}%
\frac{H_0^2\Omega_0}{2} M^{2})^{1/3}} \nonumber
\label{eq:diff}
\end{equation}
This last term changes the dependence of the temperature on
the mass, $M$, in the T-M relation. Moreover the new T-M relation
depends on the angular momentum, $L$,  
originating from the gravitational interaction of the quadrupole moment
of the protocluster with the tidal field of the matter of the
neighboring protostructures.
In Fig. 2 the X-ray temperature distribution, derived using
a CDM model with $\Omega_0=1$, $h=1/2$ and taking into account 
non-radial motions, is compared with Henry \& Arnaud (1991) and Edge et al.
(1990) data and with a pure CDM model with $\Omega_0=1$, $h=1/2$.
\begin{figure}[ht]
\psfig{file=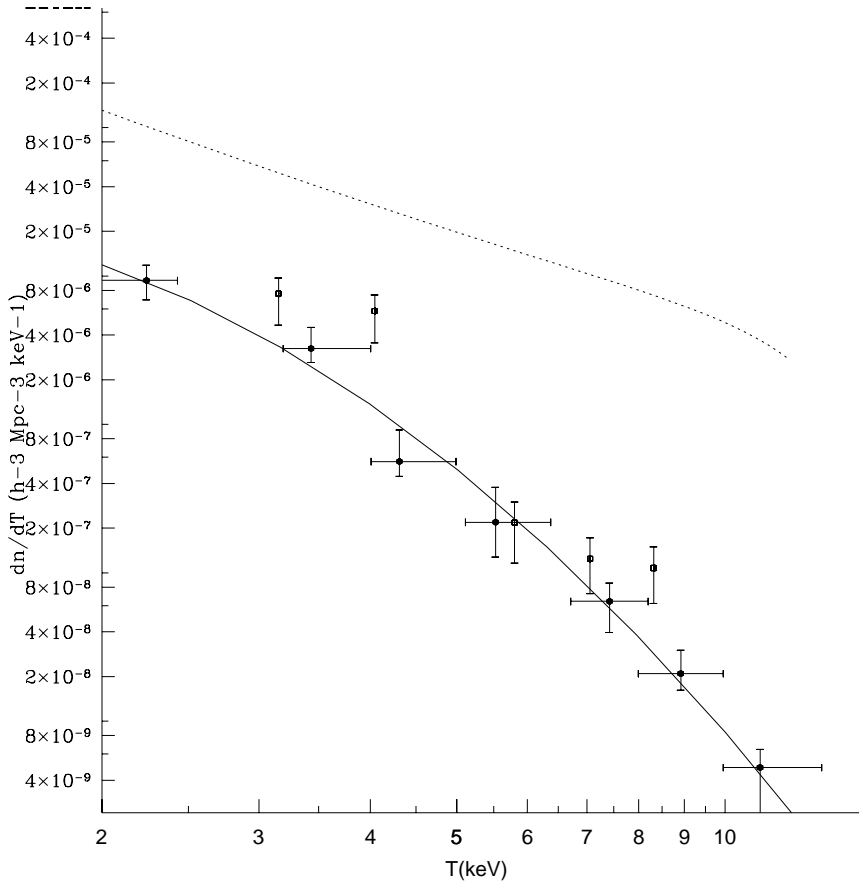,width=12cm}
\caption[]{X-ray temperature distribution function. The dashed line gives the
temperature function for a pure CDM model ($\Omega_0=1$, $h=1/2$),
with $ R_{f} = 3 h^{-1} Mpc$. The solid
line is the same distribution but now taking account of non-radial motions. 
The data are obtained by Edge et al. 1990 (dots), and Henry \& Arnaud 1991 (open squares)}
\end{figure}
As shown the CDM model that does not take account of the non-radial
motions over-produces the clusters abundance.  
The introduction of non-radial motions
gives a more careful description of the experimental data. 
As we have seen
the X-ray temperature distribution function obtained taking account
of non-radial motions is different from that of a pure CDM model for two
reasons: \\
1) the variation of the threshold, $\delta_c$, with mass, $M$.
This is due to the change  in the energetics of the
collapse model
produced by the introduction of another potential energy term
($\frac{L(r,\nu)^2}{M^2 r^3}$) in the equation decsribing the collapse
(see $Eq. (\ref{eq:coll})$);  \\
2) the modification of the T-M relation produced
by the alteration of the partition of energy in virial equilibrium. \\
For values of mass $M=0.5 M_{\odot}$ the difference between
the two theoretical lines in Fig. 2 is due to the first factor
for $\simeq 59 \%$ and this value increases with increasing mass. 
The uncertainty in our model 
fundamentally comes from the uncertainty of the T-M relation whose
value has been previously quoted.\\
Somebody may object that
the effect  here described has not been seen in some
hydrodynamic simulation
(Evrard \& Crone 1992). The answer to this objection can be given
remembering 
a similar problem of 
the previrialization conjecture 
(Davis \& Peebles 1977; Peebles 1990), (supposing that initial
asphericities and tidal interactions between neighboring density fluctuations
induce significant non-radial motions, which oppose the collapse)
on which our model is fundamentally based.
It is known that while
some N-body simulations (Villumsen \& Davis 1986; Peebles 1990)
appear to reproduce this effect, other simulations
(for example Evrard \& Crone 1992) do not. An answer to this controversy
was given by Lokas et al. 1996. The problem is connected to  the spectral
index $n$ used in the simulations. The "previrialization"
is seen only for $n>-1$. While Peebles (1990) used simulations with $n=0$,
Evrard \& Crone (1992) assumed $n=-1$.
Excluding this particular case, generally the whole  properties of
clusters, such as  their optical and X-ray luminosity functions, or their
velocity and temperature distribution functions, are difficult to address
directly in numerical simulations because the size of the box must be very
large in order to contain a sufficient number of clusters. Then  the analytical
approach remains an effective alternative.\\

\section{Non-radial motions and the clusters correlation function}
 
As previously seen in Sect. 2, in the PS theory 
the comoving number density of non-linear objects of mass $M$ to $M+dM$
is simply obtained by differentiating with respect to mass the
integral from $\delta_c$ to infinity 
of the probability distribution for fluctuations given by:
\begin{equation}
p[\delta (M)]=\frac 1{\sqrt{2\pi }\sigma (M)}\exp [-\delta (M)^2/2\sigma
(M)^2] \label{eq:pip}
\end{equation}
In an exactly analogous way, the probability $p(M_1,M_2,r)$ per unit volume
per unit masses of finding two collapsed objects of mass $M_1$ and $M_2$
separated by a distance $r$ is obtained by integrating both variables of the
bivariant Gaussian distribution in $\delta (M_1)$ and $\delta (M_2)$, with
correlation $\xi _\rho (r)$ from
$\delta _c$ to infinity and then taking the
partial derivatives with respect to both masses. The correlation function
for collapsed objects is simply obtained from:
\begin{equation}
\xi _{MM}=p(M_1,M_2,r)/p(M_1)p(M_2)-1 
\end{equation}
\noindent which for equal masses and
on scales bigger than that of the turn-around ($\xi _\rho \ll 1$) is
(Kashlinsky 1987):
\begin{equation}
\xi _{MM}=[\delta _c^2/\sigma(M)^2]\xi _\rho (r)  \label{eq:coo}
\end{equation}
where
$\xi_{\rho}$ is the correlation function of the matter density
distribution when the density fluctuations had small amplitude.
Eq. (\ref{eq:coo}) shows that the correlation of collapsed objects may be
enhanced with respect to the underlying mass fluctuations. This condition
is usually described by the bias parameter $b$ which is sometimes defined
as $[\xi_{MM}(r)/\xi_{\rho}(r)]^{1/2}=\delta_{c}/\sigma(M)$.\\
Studies of clustering on scales $\geq 10h^{-1}Mpc$ have shown that
the correlation function given in Eq. (\ref{eq:coo}) is different from
that obtained from
observations. The most compelling data are angular correlation functions for
the APM survey. These decline much less rapidly on large scales than the CDM
prediction (Maddox et al. 1990a).
This discrepancy
can be reduced taking into
account the non-radial motions that originate when a cluster reaches the
non-linear regime. In fact, the calculation  of the correlation function 
requires the specification of $\delta _c$ which is modified by non-radial 
motions. \\
The result of our calculation is showed in Fig. 3.\\
The observational data to which we compare the calculated
correlation function are those obtained by Sutherland \& Efstathiou 
(1991) from the analysis
of Huchra's et al. (1990) deep redshift survey as discussed in Geller 
\& Huchra (1988) and the data points for the APM clusters 
computed by Efstathiou et al. (1992a).
\begin{figure}
\psfig{file=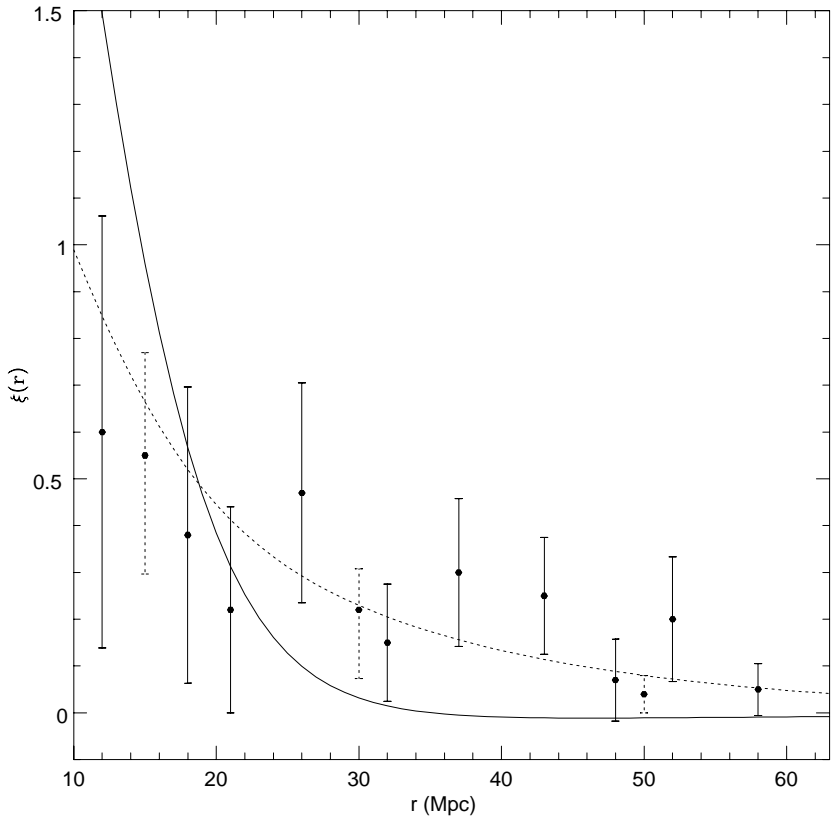,width=12cm}
\caption[]{Clusters of galaxies correlation function.
The solid line gives the
correlation function for a pure CDM model, with $R_{f}=3 h^{-1} Mpc$. The dashed
line is the same distribution but now takes account of non-radial motions. 
The observational data refer to the two-point correlation
function obtained by Sutherland \& Efstathiou (1991) ({\it filled  exagons})
from the analysis of Huchra's et al. (1990) deep redshift survey and 
with the data points for the APM clusters computed by Efstathiou et 
al. (1992a) ({\it dashed errorbars}).}
\end{figure}
The deep-cluster redshift survey of Huchra et al. (1990) consists of the
145 Abell clusters with $R \geq 0$, $D \leq 6$ in the area
$10^h \leq \alpha \leq 15^h$,
$58^0 \leq \delta \leq 78^0$. The APM sample consists of 240 clusters with APM
richness ${\it R} \geq 20$ and photometrically derived redshifts
$z_{\chi} \leq 0.1$. The survey covers a solid angle $\omega=1.3 sr$
in the region of sky $21^h \leq \alpha \leq 5^h$, $-72.5^0 \leq \delta \leq -17.5^0$ of the
APM galaxy survey.
Both in Sutherland \& Efstathiou (1991) and Efstathiou et al. (1992a) the
correlation function
was calculated from the samples using the estimator:
\begin{equation}
\xi(r)=F \frac{N_{cc}}{N_{cr}}-1
\end{equation}
where $N_{cc}$ and $N_{cr}$ are the numbers of cluster pairs and cluster-random
pairs having redshift-space separations in the range ($r-dr/2$,$r+dr/2$). 
The random
points are generated within the sample volume with a mean density F times that
of the clusters and with a redshift distribution derived from a smoothed
distribution of the redshifts for the cluster sample. The correlation
function obtained was finally
corrected 
for
line-of-sight anisotropies.
This correction is necessary because the clustering of Abell clusters
is highly anisotropic in redshift space, providing evidence that the Abell
catalogues, which were built by scanning photographic plates by eye, are
affected by incompleteness on the plane of the sky which enhances the clustering
amplitude measured in three dimensions (Sutherland \& Efstathiou 1991;
Efstathiou et al. 1992a). In particular Efstathiou et al. (1992a)
presented
a clear example of incompleteness in the Abell $R \geq 0$ catalogue
comparing the
machine-based APM survey with the Postman, Hucra \& Geller 1992 redshift
survey of Abell clusters showing that even if the clusters in these surveys
have comparable space densities the red-shift space correlation
function of the APM sample is isotropic on large scales while the correlation
function for the Abell clusters is highly anisotropic. 
In any case, after this correction the correlation function of Abell clusters
agrees extremely well with that of APM clusters (Efstathiou et al. 1992a).
We use two samples because: \\
a) we wanted to compare our model for
the two-point correlation function with observational data and at the same 
time with the result obtained by Borgani (1990) who studied
the effect of particular thresholds (erfc-threshold and Gaussian-threshold)
on the correlation properties of clusters of galaxies and compared his
result with that by Sutherland \& Efstathiou (1991). For this reason we used
this last sample; \\
2) we used the APM galaxy survey
because the uniformity of the APM magnitudes, the low obscuration in the
APM survey area and the use, for its construction, of a computer
cluster-finding algorithm (see Maddox et al. 1990a,b) further reduce
the possibility of spurious clustering on the plane of the sky. \\
The comparison between the correlation function
and
the quoted data, displayed in Fig. 3, shows that
there is an evident discrepancy between pure
CDM previsions and experimental data. The CDM model seems to have
trouble in re-producing the behaviour of the data. In fact, the predicted
two-point cluster function is too steep and rapidly goes nearly to zero for
$r \simeq 30 h^{-1} Mpc$, while the data show no significant anticorrelation
up to   $r \simeq 60 h^{-1} Mpc$ (see Borgani 1990).
This is a direct effect due to the non-scale invariance of the CDM spectrum
(Primack \& Blumenthal 1983).
The introduction of non-radial motions
gives a more accurate description of the experimental data showing that
physical effects cannot be ignored in the study of the formation of
cosmic structures.
The result obtained is in agreement with that obtained by Borgani (1990)
who showed how the introduction of smooth thresholds (similar to that
obtained in our model)
leads to two-point correlation functions in a systematically agreement with
the data.
Before going on we want to remember that the threshold functions are strictly 
connected to the concept of bias.
In fact, according to the biased theory of galaxy formation, observable 
objects of mass $\simeq M$
arise from fluctuations of the density field, filtered on a scale
$R$, 
rising over a {\it global} threshold, $\delta>\delta_c=\nu_t \sigma$, where
$\sigma$ is the rms value of $\delta$ and $\nu_t$ is the threshold height.
The number density of objects,$n_{pk}$, that forms from peaks of density
of height $\nu$ can
be written following BBKS (1986) in the form:
\begin{equation}
n_{pk}=\int_{0}^{\infty} t(\frac{\nu}{\nu_t}) N_{pk}(\nu) d\nu
\end{equation}
where $t(\frac{\nu}{\nu_t})$ is the threshold function, $\nu_t$ the threshold
height and $N_{pk}d\nu$ the differential number density of peaks
(see BBKS 1986 Eq. 4.3). The threshold level $\nu_t$ is defined so
that the probability of a peak becoming an observable object is $1/2$ when
$\nu=\nu_t$. In the sharp threshold case the selection
function, is a Heaviside function
$t(\frac{\nu}{\nu_t})=\theta(\nu-\nu_t)$. 
As previously quoted the threshold function is connected to the bias 
coefficient of a class of objects by (BBKS): 
\begin{equation}
b(R_f)= \frac{<\tilde\nu> }{\sigma_o}+1
\end{equation}
where $< \tilde\nu >$ is: 
\begin{equation}
< \tilde\nu> = \int_0^\infty \left[ \nu -\frac{\gamma \theta }{%
1-\gamma ^2}\right] t(\frac{\nu}{\nu_{t}}) N_{pk}(\nu ) d\nu  \label{eq:nu}
\end{equation}
while, $\gamma $ and $\vartheta $ are given in BBKS (1986)
(respectively Eq. 4.6 a; Eq. 6.14).\\
While in a $\theta$ threshold scheme, fluctuations below $\delta_c$ have zero
probability to develop an observable object and fluctuations above $\delta_c$
have zero probability not to develop an object, the situation is totally
different when an erfc-threshold, as that introduced by Borgani (1990),
is used. In this case objects can also be formed from fluctuations
below $\delta_c$ and
there is a non-zero probability for fluctuations above $\delta_c$ to
be sterile. \\
According with Borgani (1990), the erfc-threshold
can be related to non-sphericity effects during the gravitational growing
process. In fact, while in the spherical limit it is possible to find a precise
relation between the time, $t$ elapsed from the turning around and the
correspondent density contrast, $\delta$,($t \propto \delta^{-3/2})$
(Gunn \& Gott 1972), when non-sphericity is introduced it is no longer
possible to univocally relate the primeval density contrast and the
evolutionary stage of a fluctuation 
because while in the spherical model an object is characterized
by $\delta>\delta_c$
the non-sphericity produces a spread around the typical value $\delta_c$.
The erfc-threshold is a way to correct the quoted $t-\delta$ relation.
The erfc-threshold is also linked to non-radial motions.
In fact as we previously told, the tidal interaction of
the irregular mass distribution within and around the
protocluster, present in hierarchical models, gives rise to non-radial
motions. We also told that the erfc-threshold is related to
non-sphericity effects that
in turn are responsible for the origin of the quoted non-radial motions. \\
Borgani's (1990) model, like our,  is characterized by a non-$\theta$
threshold. As we showed in a previous paper, Del Popolo \& Gambera 1998a,
one of the effects of non-radial motions is that the threshold function   
differs from a Heaviside function (sharp threshold),
(see Fig. 7 of Del Popolo \& Gambera 1998a).
In this last paper the threshold function is defined as:
\begin{equation}
t(\nu )=\int_{\delta _c}^\infty p\left[ \overline{\delta} ,
\langle \overline{\delta} (r_{Mt},\nu
 )\rangle ,\sigma _{\overline{\delta}} (r_{Mt},\nu )\right] d\delta \label{eq:sel}
\end{equation}
where the function 
\begin{equation}
p\left[ \overline{\delta} ,\langle \overline{\delta} (r) \rangle\right] = 
\frac 1{\sqrt{2\pi }\sigma_{\overline{\delta}} }\exp \left( -\frac{|\overline{\delta} -
\langle \overline{\delta} (r) \rangle|^2}{2\sigma
_{\overline{\delta}} ^2}\right) \label{eq:gau}
\end{equation}
gives the probability that the peak overdensity is
different from the average, in a Gaussian density field.
As displayed, the integrand is evaluated at a radius $r_{Mt}$ which is the
typical radius of the object that we are selecting. Moreover, the threshold 
function $t(\nu )$ depends on the critical overdensity threshold for the
collapse, $\delta _{c}$, which is not constant as in a spherical model
(due to the presence, in our analysis, of non-radial motions that delay the
collapse of the proto-cluster) (see Eq. (\ref{eq:ma7})).
The fundamental difference between our and Borgani's approach is that our
threshold function is physically motivated: it is simply obtained from the
assumptions of a Gaussian density field and taking account of non-radial
motions. Borgani's threshold functions (erfc and Gaussian threshold) are
ad-hoc introduced in order to reduce the discrepancy between the observed and
the CDM predicted two-point correlation functions of clusters of galaxies.
The connection with the quoted non-sphericity effects, even if
logical and in agreement with our results, is only a posteriori tentative
to justify the choice made.

\section{Conclusions}

                           
In these last years many authors (Bahcall \& Soneira 1983; 
White et al. 1987; Maddox et al. 1990a; 
Saunders et al. 1991; Peacock 1991; Bartlett \& Silk 1993) have shown the existence of a 
strong discrepancy between the 
observed X-ray temperature distribution function 
of clusters and that predicted by a CDM model
and the observed two-point correlation function of
clusters and that predicted by the CDM model. 
To reduce these discrepancies several alternative models have been introduced 
but no model has considered the role of the non-radial motions. Here 
we have shown how non-radial motions may reduce both the two quoted 
discrepancies.
To this aim we calculated
the variation in the treshold parameter,
$\delta_{c}$, as a function of the mass $M$, 
and that of the temperature-mass relation, 
produced by the presence
of non-radial motions in the outskirts of clusters of galaxies.
We compared
Edge et al. (1990) and Henry \& Arnaud (1991) data with 
the distribution function of X-ray temperature, 
calculated using
PS (1974) theory and 
Evrard's (1990) prescriptions for the mass-temperature 
relation. 
We found that the model produces a reasonable 
clusters temperature distribution (see Fig. 2).
We also used $\delta_{c}(M)$ to calculate the two-point correlation of
clusters of galaxies and we compared it both with
that obtained
by Sutherland \& Efstathiou (1991) from the analysis
of Huchra's et al. (1990) deep redshift survey as  
discussed in Geller \& Huchra (1988) and with the
data points for the APM clusters computed by Efstathiou et al. (1992a). 
Our results (see Fig. 3) show how non-radial motions change the
correlation length of the correlation function making it less steep 
than that obtained from a pure CDM model where the non-radial motions are 
not considered. 
\begin{flushleft}
{\it Acknowledgements}
\end{flushleft}
 We are grateful to E. Spedicato and to V. Antonuccio-Delogu for 
stimulating discussions during the period in which this work was performed. 
We thank the anonymous referees whose comments and suggestions help 
us to improve the quality of this work.\\


\begin{thebibliography}{}
\bibitem{} Antonuccio-Delogu, V., Colafrancesco, S., 1995,  {\it preprint Sissa astro-ph/9507005}
\bibitem{} Babul A., White S.D.M., 1991, MNRAS, 253, 31P
\bibitem{} Bahcall, N.A., Soneira, R.M., 1983, ApJ 270, 20
\bibitem{} Bardeen J.M., Bond J.R., Kaiser N., Szalay A.S., 1986, ApJ, 304, 15 (BBKS)
\bibitem{} Barrow, J.D., Silk, J., 1981, ApJ 250, 432
\bibitem{} Bartlett J.G., 1997, Sissa preprint, astro-ph/9703090
\bibitem{} Bartlett, J.G., Silk, J., 1993, ApJ 407, L45
\bibitem{} Blumenthal, G.R., Faber S.G., Primack, J.R, Rees, M.J., 1984, Nat., 311, 517
\bibitem{} Bond J.R., Cole S., Efstathiou G., Kaiser N., 1991, ApJ, 379, 440
\bibitem{} Bond J.R., Efstathiou G., 1984, ApJ, 285, L45 
\bibitem{} Bond, J.R., Szalay A.S., Silk J., 1988, ApJ, 324, 627
\bibitem{} Borgani, S., 1990, A\&A 240, 223
\bibitem{} Bower R.G., Coles P., Frenk C.S., White S.D.M., 1993, ApJ, 405, 403
\bibitem{} Brainerd T.G., Villumsen J.V., 1992, ApJ, 394, 409
\bibitem{} Cavaliere A., Colafrancesco S., 1988, ApJ, 331, 660
\bibitem{} Cen, R.Y., Gnedin, N.Y., Kofman, L.A., Ostriker, J.P., 1992 preprint
\bibitem{} Crone, M.M., Evrard A.E., Richstone, D.O., 1994, ApJ 434, 402
\bibitem{} Dalton, G. B., Efststhiou, G., Maddox, S. J., Sutherland, W. J., 1992, ApJ 390, L1
\bibitem{} Davis M., Peebles P.J.E., 1977, ApJS., 34, 425 
\bibitem{} Davis M., Summers F.J., Schlegel D., 1992, Nat., 359, 393
\bibitem{} Dekel A., Bertshinger E., Yahil A. et al. 1992,  IRAS galaxies verses POTENT mass: density fields, biasing and $ \Omega$, Princeton preprint IASSNS-AST 92/55
\bibitem{} Dekel, A., Aarseth, S.J., 1984, ApJ 238, 1
\bibitem{} Dekel, A., Blumenthal, G.R., Primack, J.R., Olivier, S., 1989, ApJ 338, L5
\bibitem{} Del Popolo, A., Gambera, M., 1998a, A \& A accepted (see SISSA preprint astro-ph/9802214
\bibitem{} Del Popolo, A., Gambera, M., 1998b, Hadronic Jour. Suppl.,  {\it accepted} 
\bibitem{} Del Popolo, A., Gambera, M., 1998c, A\&A, {\it submitted} 
\bibitem{} Edge A.C., Stewart G.C., Fabian A.C., Arnaud K.A., 1990, MNRAS, 245, 559
\bibitem{} Efstathiou G., 1990, in ''The physics of the early Universe'', eds Heavens A., Peacock J., Davies A., (SUSSP)
\bibitem{} Efstathiou G., Bond J.R., White S.D.M., 1992, MNRAS, 258, 1P
\bibitem{} Efstathiou G., Frenk C.S., White S.D.M.., Davis M., 1988, MNRAS, 235, 715
\bibitem{} Efstathiou G., Rees M.J., 1988, MNRAS, 230, 5p
\bibitem{} Efstathiou G., Sutherland W.J., Maddox,S.J., 1990a, Nat., 348, 705
\bibitem{} Efstathiou, G., Dalton, G. B., Sutherland, W. J., 
Maddox, S. J., 1992a, MNRAS 257, 125
\bibitem{} Efstathiou, G., Kaiser, N., Saunders, W. et al. 1990b, MNRAS 247, 10p
\bibitem{} Evrard A.E., 1989, ApJ, 341, L71
\bibitem{} Evrard A.E., 1990, in Proc. STScI Symp.4, ed. W.R. Oegerle, M.J. Fitchett, \& L. Danly (New York: Cambridge Univ. Press), 287
\bibitem{} Evrard A.E., 1991, in Clusters of Galaxies ed. M. Fitchett \& W. Oegerle (Cambridge: Cambridge Univ. Press.)
\bibitem{} Evrard A.E., Crone M.M., 1992, ApJ, 394, L1
\bibitem{} Evrard, A. E., 1997, Sissa Preprint astro-ph/9701148
\bibitem{} Evrard, A. E., Metzler, C. A., Navarro, J. F., 1996, ApJ 469, 494
\bibitem{} Filmore, J. A., Goldreich, P., 1984, ApJ 281, 1
\bibitem{} Frenk C.S., White S.D.M., Davis M., Efstathiou G., 1988 ApJ, 327, 507
\bibitem{} Geller, M.J., Huchra, J.P., 1988, in Large Scale Motions in the Universe, eds. V. C. Rubin, G.V., Coyne, Princeton University Press, Princeton
\bibitem{} Gunn J.E., Gott J.R., 1972, ApJ, 176, 1
\bibitem{} Gunn, J.E., 1977, ApJ 218, 592
\bibitem{} Hauser, M.G., Peebles, P.J.E., 1973, ApJ 185, 757 
\bibitem{} Henry J.P., Arnaud K.A., 1991, ApJ, 372,410
\bibitem{} Hoffman, Y., Shaham, J., 1985, ApJ 297, 16 
\bibitem{} Holtzman J., 1989, ApJS, 71, 1
\bibitem{} Holtzman J., Primack J., 1993, Phys. Rev. D43, 3155
\bibitem{} Huchra, J.P.,  Henry, J.P., Postman, M., Geller, M.J., 1990, ApJ 365, 66
\bibitem{} Kaiser N., Efstathiou G., Ellis R. et al. 1991, MNRAS, 252, 1
\bibitem{} Kaiser N., Lahav O., 1989, MNRAS, 1989, 237, 129
\bibitem{} Kaiser, N., 1984, Ap. J., 284, L 9
\bibitem{} Kashlinsky, A., 1987, ApJ 317, 19
\bibitem{} Klypin, A.A., Kopylov, A.I., 1983, Soviet Astron. Lett., 9, 41
\bibitem{} Klypin, A. A., Nolthenius R., Primack J., 1997, ApJ, 474, 533
\bibitem{} Lacey C., Cole S., 1994, MNRAS, 271, 676
\bibitem{} Lahav O., Edge A., Fabian A.C., Putney A., 1989, MNRAS, 238, 881
\bibitem{} Lahav O., Rowan-Robinson M., Lynden-Bell D., 1988, MNRAS, 234, 677
\bibitem{} Liddle A.R., Lyth D.H., 1993, Phys. Rep., 231, n 1, 2
\bibitem{} Lokas E.L., Juskiewicz R., Bouchet F.R., Hivon E., 1996, ApJ, 467, 1
\bibitem{} Maddox S.J., Efstathiou G., .Sutherland W.J., Loveday J., 1990a, MNRAS 242, 43p
\bibitem{} Maddox S.J., Efstathiou, G., Sutherland, W.J., Loveday, J., 1990b, MNRAS 243, 692
\bibitem{} Maddox S.J., Efstathiou, G., Sutherland, W.J., MNRAS 246, 433
\bibitem{} Olivier, S., Blumenthal, G.R., Primack, J.R., Stanhill, D., 1990, ApJ 356, 1
\bibitem{} Oukbir J., Bartlett J.G., Blanchard A., 1996, Sissa preprint astro-ph/9611089
\bibitem{} Peacock J.A., 1991, MNRAS, 253, 1p
\bibitem{} Peacock J.A., Heavens A.F., 1990, MNRAS, 243, 133
\bibitem{} Peacock J.A., Nicholson D., 1991, MNRAS, 253, 307
\bibitem{} Peebles P. J. E., 1984, Ap. J., 284, 439
\bibitem{} Peebles P.J.E., 1982, ApJ, 258, 415
\bibitem{} Peebles P.J.E., 1990, ApJ, 365, 27
\bibitem{} Peebles P.J.E., 1980, The large scale structure of the Universe, Princeton university press
\bibitem{} Peebles, P.J.E., 1993, Principles of Physical Cosmology, Princeton University Press
\bibitem{} Pen U., 1997, Sissa preprint astro-ph/9610147
\bibitem{} Postman, M., Huchra, J. P., Geller, M. J., 1992, ApJ, 384, 404
\bibitem{} Press W.H., Schechter P., 1974, ApJ, 187, 425 (PS)
\bibitem{} Primack, S. M., Blumenthal, G. R., 1983, in Formation and Evolution of Galaxies and Large Structures in The Universe, eds. H.A. Weldon, P. Langacker, P. J. Steinhardt, Birkh\"auser, Boston.
\bibitem{} Quinn, P.J., Salmon, J.K., Zurek, W.H., 1986, Nature, 322, 329
\bibitem{} Quinn, P.J., Zurek, W.H., 1988, ApJ 331, 1
\bibitem{} Ryden B.S., Gunn J.E., 1987, ApJ, 318, 15
\bibitem{} Ryden, B.S., 1988, ApJ 329, 589
\bibitem{} Saunders, W., et al., 1991, Nature 349, 32 
\bibitem{} Schaefer R.K., 1991, Int. J. Mod. Phys., A6, 2075
\bibitem{} Schaefer R.K., Shafi Q., 1993, Phys. Rev., D47, 1333
\bibitem{} Schaefer, R.K., Shafi, Q., Stecker, F., 1989, ApJ 347, 575
\bibitem{} Shafi, Q., Stecker, F.W., 1984, Phys. Rev. D 29, 187
\bibitem{} Smoot G.F., et al., 1992, ApJ, 396, L1
\bibitem{} Sutherland W.J., 1988, MNRAS 234, 159
\bibitem{} Sutherland, W., Efstathiou, G., 1991, MNRAS 248, 159
\bibitem{} Szalay, A.S., Silk, J., 1983, ApJ 264, L31
\bibitem{} Turner, M.S., 1991, Phys. Scr. 36, 167
\bibitem{} Valdarnini, R., Bonometto, S.A., 1985, A\&A 146, 235
\bibitem{} Villumsen, J.V., Davis, M., 1986, ApJ, 308. 499
\bibitem{} Warren, M. S., Zurek, W. H., Quinn, P. J., Salmon, J. K., 1991, in After the First Three minutes, ed. S. Holt, V. Trimble, \& C. Bennett (New York: AIP), 210
\bibitem{} White S.D.M., Efstathiou G., Frenk C.S., 1993a, The amplitude of the mass fluctuations in the Universe, Durham preprint
\bibitem{} White S.D.M.., Efstathiou G., Frenk C.S., 1993b, MNRAS, 262, 1023
\bibitem{} White S.D.M.., Frenk C.S., Davis M., Efstathiou G., 1987, ApJ, 313, 505
\bibitem{} White, S.D.M., Davis, M., Efstathiou, G., Frenk C.S., 1987, Nature 330, 451
\bibitem{} Yano T., Masahiro N., Gouda N., 1996, ApJ, 466, 1
\bibitem{} Zurek, W.H., Quinn, P.J., Salmon, J.K., 1988, ApJ 330, 519
\end{thebibliography}
\end{document}